# Noise-immune cavity-enhanced optical frequency comb spectroscopy: A sensitive technique for high-resolution broadband molecular detection

Amir Khodabakhsh, Alexandra C. Johansson, and Aleksandra Foltynowicz[*]

*Department of Physics, Umeå University, 901 87 Umeå, Sweden*

*email: aleksandra.foltynowicz@physics.umu.se*

Noise-immune cavity-enhanced optical frequency comb spectroscopy (NICE-OFCS) is a recently developed technique that utilizes phase modulation to obtain immunity to frequency-to-amplitude noise conversion by the cavity modes and yields high absorption sensitivity over a broad spectral range. We describe the principles of the technique and discuss possible comb-cavity matching solutions. We present a theoretical description of NICE-OFCS signals detected with a Fourier transform spectrometer (FTS), and validate the model by comparing it to experimental $CO_2$ spectra around 1575 nm. Our system is based on an Er:fiber femtosecond laser locked to a cavity and phase-modulated at a frequency equal to a multiple of the cavity free spectral range (FSR). The NICE-OFCS signal is detected by a fast-scanning FTS equipped with a high-bandwidth commercial detector. We demonstrate a simple method of passive locking of the modulation frequency to the cavity FSR that significantly improves the long term stability of the system, allowing averaging times on the order of minutes. Using a cavity with a finesse of ~9000 we obtain absorption sensitivity of $6.4 \times 10^{-11}$ cm$^{-1}$ Hz$^{-1/2}$ per spectral element, and concentration detection limit for $CO_2$ of 450 ppb Hz$^{-1/2}$, determined by multiline fitting.

## 1. Introduction

Cavity-enhanced optical frequency comb spectroscopy (CE-OFCS) [1, 2] has been gaining increased attention in the research community during the last decade because of the unique combination of broad spectral bandwidth, high resolution and absorption sensitivity that it provides. The technique allows detection of entire molecular absorption bands in acquisition times down to μs [3, 4] and enables simultaneous multispecies detection with high species selectivity [2, 5, 6]. Until now, the potential of CE-OFCS has been demonstrated for applications ranging from breath analysis [5, 6], time-resolved chemical sensing [4], flame characterization [7], industrial process control [8], to environmental monitoring [9].

In order to take advantage of the benefits offered by optical frequency combs (OFC) for spectroscopy, various detection schemes have been developed capable of analyzing the broad spectral information carried by an OFC [10-13]. In general, they are based either on a dispersive element in combination with a detector, detector array or a camera [1, 5, 14-17], or on Fourier transform spectrometry (FTS) employing either a mechanical interferometer [18, 19] or the dual-comb approach [3, 20, 21]. Of these, a mechanical FTS provides the advantage of acquiring the entire spectral bandwidth of an OFC with high resolution using a single detector in acquisition times on the order of a second. This approach is easily implemented in both the near- [18, 19] and mid-infrared [6, 22] wavelength ranges, and it is compatible with commercial FTIR instruments [23, 24].

The broad equidistant spectrum of an optical frequency comb can be efficiently coupled to an external cavity by matching the repetition rate, $f_{rep}$, of the comb to the cavity free spectral range (FSR) and adjusting the carrier-envelope offset frequency, $f_{ceo}$, so that the comb lines come into resonance with the corresponding cavity modes. This matching has to be actively sustained in order to compensate for the drift of cavity length and/or comb parameters. One way of coupling the comb to the cavity is to dither the cavity length or comb lines around the perfect matching condition, so that the comb comes periodically into resonance with the cavity. This approach is often used in combination with detection schemes based on dispersion elements, in which the cavity transmission is averaged over many dither cycles [5, 16]. However, it is not compatible with a fast-scanning FTS, in which an interferogram is acquired at a sampling rate (~Msamples/s) higher than the possible dither frequency (usually in the kHz range, limited by the bandwidth of comb or cavity actuators). This precludes efficient averaging of the transmitted comb intensity and causes, in turn, distortion in the interferogram and the final absorption spectrum. Thus a tight lock of the comb to the cavity is required when a fast-scanning FTS is used as a detection method [6, 19]. The tight comb-cavity locking introduces amplitude noise in the transmitted spectrum, originating from the residual fluctuations of the comb line frequencies relative to the center frequencies of the narrow cavity modes; a process called frequency-to-amplitude noise conversion. This noise can be reduced down to the shot-noise-level using an auto-balancing detector to subtract the signals from the two outputs of an interferometer [19]. However, this approach requires a custom-built detector, and the achieved noise reduction depends critically on the performance of the detector.

We have recently shown [25] that noise in the FTS-based CE-OFCS can instead be reduced in the optical domain by employing a concept known from continuous wave (cw) laser-based absorption spectroscopy and utilized in noise-immune cavity-enhanced optical heterodyne molecular spectroscopy (NICE-OHMS) [26]. In NICE-OHMS, the laser light is phase-modulated at a frequency equal to the FSR of an external cavity, so that when the laser carrier frequency is locked to a cavity mode, its sidebands are transmitted through the adjacent cavity modes. As in ordinary frequency modulation spectroscopy (FMS) [27], the transmitted intensity contains a beat signal at the modulation frequency whose amplitude is proportional to (depending on the detection phase) the difference of the phase shift of the carrier and the mean value of the phase shift of the sidebands, or the difference of the attenuations of the sidebands. When the modulation frequency is perfectly matched to the cavity FSR, the carrier and the sidebands are transmitted through their corresponding cavity modes in exactly the same way and any residual frequency noise on the laser affects them in an identical manner. Thus no signal is detected at the modulation frequency and the technique gains immunity to frequency-to-amplitude noise conversion by the cavity.

We implemented this concept in near-infrared CE-OFCS by locking an Er:fiber OFC to an external cavity, phase-modulating it at a frequency equal to the cavity FSR and detecting the transmitted intensity with a fast-scanning FTS equipped with a high-bandwidth detector. In the first demonstration of noise-immune cavity-enhanced optical frequency comb spectroscopy (NICE-OFCS) [25] we obtained close to shot-noise-limited absorption sensitivity of $4.3 \times 10^{-10}$ cm$^{-1}$ Hz$^{-1/2}$ per spectral element in a cavity with a finesse of ~2000. Here we present the NICE-OFCS method in more detail, we discuss various comb-cavity matching solutions, and describe the theoretical model of the signals. Compared to the first demonstration, we increase the detection sensitivity by using a cavity with a higher finesse and we improve the long term stability of the system by locking the modulation frequency to the cavity FSR. We show the validity of the theoretical model by comparing it to experimental NICE-OFCS signals from the overtone $CO_2$ band at 1575 nm. We use multiline fitting and show that it improves the concentration detection limits compared to those obtained from the signal-to-noise ratio of a single absorption line.

## 2. Comb-cavity matching in NICE-OFCS

In NICE-OFCS the frequency comb is locked to a cavity and phase-modulated at a frequency equal to (a multiple of) the cavity FSR. Each comb line and sideband needs to be transmitted through a separate cavity mode in order to prevent overlap and unwanted interference between the various frequency components. To transmit all comb lines and their sidebands through the cavity, the ratio between the laser $f_{rep}$ and the cavity FSR has to be at least 3, so that there are at least two cavity modes in between those to which the comb lines are locked. The conceptually simplest way to achieve this is shown in Fig. 1(a), where each triplet is indicated by a different color. Note that in this configuration the modulation frequency, $f_m$, can be equal to the FSR (as shown in the figure) as well as to $(3m\pm1)$FSR, where $m$ is an integer.

Since the typical repetition rate of commercially available visible and near-infrared comb sources is in the 80-250 MHz range, a linear cavity with FSR = $f_{rep}$/3 becomes impractically long. For example, the $f_{rep}$ of our Er:fiber comb is equal to 250 MHz yielding a cavity FSR of 83.3 MHz, corresponding to a length of 1.8 m. To decrease the cavity length other ratios of FSR and $f_{rep}$ must be used, however, at the expense of losing some of the OFC lines in the transmission.

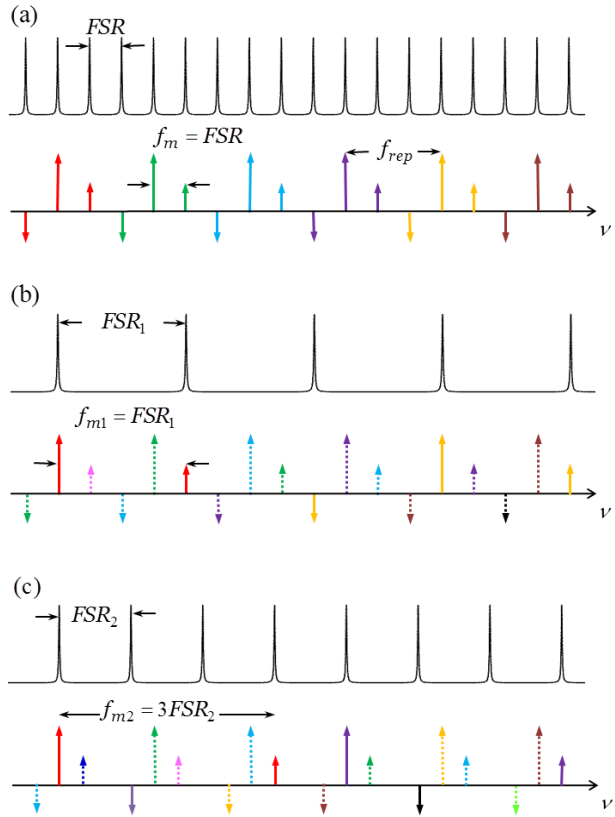

Fig. 1. Examples of comb-cavity matching schemes for implementing NICE-OFCS. The transmitted and reflected modes and sidebands are shown by solid and dashed lines, respectively. (a) Matching that allows transmitting all incident comb lines. (b) Matching used in the initial demonstration that yields a shorter cavity length. (c) Matching implemented in this work that allows transmitting more of the incident comb power than (b) and removes the beating of the sidebands of different comb lines at the modulation frequency.

The comb-cavity matching solution that we used in the initial demonstration of NICE-OFCS [25] is shown in Fig. 1(b). In this case every 4$^{th}$ OFC line is matched to every 3$^{rd}$ cavity mode, and the modulation frequency should again be equal to the cavity FSR or to $(3m\pm1)$FSR. Thus every 4$^{th}$ comb line and its sidebands are transmitted through the cavity, which for our comb had a compact length of 45 cm. A disadvantage of this approach is that 75% of the OFC power is reflected and lost for the spectroscopic measurement, and that the sidebands of two different comb lines beat with each other at the modulation frequency, potentially introducing drift and distortion to the NICE-OFCS spectrum.

In order to increase the fraction of the transmitted power and remove the beating of the sidebands belonging to different comb lines while keeping the cavity length reasonable, we introduce in this work a different comb-cavity matching solution that is shown in Fig. 1(c). In this solution every 3$^{rd}$ comb line is matched to every 4$^{th}$ cavity mode, so that 33% of the incident comb lines are transmitted, and the modulation frequency should be equal to the cavity FSR or $(4m\pm1)$FSR (the particular case of $f_{m2} = 3\text{FSR}_2$ is shown in the figure). The cavity length is in our case increased to 80 cm, which also has the advantage of increasing the interaction length between the comb light and the sample compared to the previous cavity.

## 3. Theory of FTS-based NICE-OFCS signals

In NICE-OFCS the phase-modulated OFC light that is transmitted through the cavity is analyzed by an FTS, where it is split into two interferometer arms, reflected from a moving mirror (and thus Doppler-shifted), and finally recombined at the output of the FTS to yield an interferogram. When the OFC field is phase-modulated sinusoidally with a small modulation index, $\beta < 1$, one pair of sidebands, separated from the carrier by the modulation frequency, $\omega_m = 2\pi f_m$, will be created for each comb line [27]. In the case when both mirrors in the FTS are moving (in opposite directions), the two phase-modulated and Doppler-shifted electric fields, $E_\pm$, at the output of the FTS are given by [25]

$$E_\pm = \sum_n \sum_{k=-1,0,1} \frac{E_n}{4} J_k(\beta) T_{n,k} e^{i\left[(\omega_n + k\omega_m)\left(t \pm \frac{\Delta}{2c}\right)\right]} + c.c., \quad (1)$$

where $E_n$ is the field amplitude and $\omega_n = 2\pi(nf_{rep} + f_{ceo})$ is the angular frequency of the $n^{th}$ comb line, $J_k(\beta)$ is the Bessel function of order $k$, $\Delta$ is the optical path difference (OPD) between the arms in the FTS, $c$ is the speed of light in vacuum and $T_{n,k} = T(\omega_n \pm k\omega_m)$ is the complex transmission function of the cavity containing the analyte for the $n^{th}$ comb line $(k = 0)$ or its sidebands $(k = \pm 1)$, given by [6]

$$T_{n,k} = \frac{t e^{-i\varphi_{n,k}/2 - \delta_{n,k} - i\phi_{n,k}}}{1 - r e^{-i\varphi_{n,k} - 2\delta_{n,k} - 2i\phi_{n,k}}}, \quad (2)$$

Oct 31, 2014

where $t$ and $r$ are the frequency-dependent intensity transmission and reflection coefficients of the cavity mirrors, respectively, and $\varphi_{n,k} = 2(\omega_n \pm k\omega_m)n_r L/c$ is the round-trip phase shift caused by the cavity, where in turn $n_r$ is the index of refraction inside the cavity and $L$ is the cavity length. The single-pass amplitude attenuation and phase shift of the light caused by the analyte are given by

$$\delta_{n,k} = \frac{1}{2} S c_{rel} pL \, \text{Re} \, \chi_{n,k} \tag{3}$$

and

$$\phi_{n,k} = \frac{1}{2} S c_{rel} pL \, \text{Im} \, \chi_{n,k}, \tag{4}$$

respectively, where $S$ is the linestrength [cm$^{-2}$/atm], $c_{rel}$ is the relative concentration of the analyte, $p$ is the pressure [atm], and $\chi_{n,k}$ is the complex lineshape function [cm].

The intensity of the light at the output of the FTS contains a term oscillating at the modulation frequency, which has two contributions. The first contribution is the beating of the comb lines of the field travelling in each arm with their own sidebands, which after demodulation at $f_m$ yields a DC offset. The second contribution is the beating of the comb lines of the electric field traveling in one arm with the sidebands of the field from the other arm (and vice versa), which constitutes the NICE-OFCS interferogram. The NICE-OFCS interferogram can be written as

$$\begin{aligned}
I^{NICE-OFCS}_{\omega_m} = & \\
& J_0(\beta) J_1(\beta) \sin(\omega_m t) \times \\
& \sum_n I_n \left\{ \cos\left(\omega_m \frac{\Delta}{2c}\right) \cos\left(\omega_n \frac{\Delta}{c}\right) \text{Im}\left(T_{n,0} T^*_{n,-1} - T^*_{n,0} T_{n,+1}\right) \right. \\
& \left. + \sin\left(\omega_m \frac{\Delta}{2c}\right) \sin\left(\omega_n \frac{\Delta}{c}\right) \text{Im}\left(T_{n,0} T^*_{n,-1} + T^*_{n,0} T_{n,+1}\right) \right\} \\
& - J_0(\beta) J_1(\beta) \cos(\omega_m t) \times \\
& \sum_n I_n \left\{ \cos\left(\omega_m \frac{\Delta}{2c}\right) \cos\left(\omega_n \frac{\Delta}{c}\right) \text{Re}\left(T_{n,0} T^*_{n,-1} - T^*_{n,0} T_{n,+1}\right) \right. \\
& \left. + \sin\left(\omega_m \frac{\Delta}{2c}\right) \sin\left(\omega_n \frac{\Delta}{c}\right) \text{Re}\left(T_{n,0} T^*_{n,-1} + T^*_{n,0} T_{n,+1}\right) \right\},
\end{aligned} \tag{5}$$

where the comb line intensity $I_n = c\varepsilon_0 E_n^2/2$ has been introduced. This signal contains one in-phase [$\sin(\omega_m t)$], and one out-of-phase [$\cos(\omega_m t)$] component. Each of these comprises, in turn, a sum of two terms consisting of a product of three factors. The first factor is the sine/cosine function of the down-converted modulation frequency, $\omega_m \Delta / 2c$, which gives rise to a slowly varying envelope over the interferogram. The second factor is the sine/cosine function of the down-converted optical frequency, $\omega_n \Delta / c$, which, after summation over all frequencies, constitutes the quickly oscillating part of the interferogram. The last factor includes the contributions from the complex cavity transmission function for each comb line and its sidebands.

The full formula above must be used to model the signals measured for arbitrary absorption values. However, to show the functional form of the signal in a more intuitive way it is useful to express it in the weakly absorbing sample approximation, where both $|\delta_{n,0} - \delta_{n,\pm 1}|$ and $|\phi_{n,0} - \phi_{n,\pm 1}|$ are $\ll 1$. For small molecular attenuation and phase shift the cavity transmission function can be approximated by $T_{n,k} = 1 - 2F(\delta_{n,k} + i\phi_{n,k})/\pi$, where $F$ is the frequency-dependent cavity finesse and where the cavity phase shift is assumed to be a multiple of $2\pi$. Under these conditions, and neglecting the terms proportional to $\delta_{n,k}^2$ and $\phi_{n,k}^2$, Eq. (5) can be rewritten as

$$
\begin{aligned}
I_{\omega_m}^{NICE-OFCS} = \\
J_0(\beta) J_1(\beta) \sin(\omega_m t) \times \\
\sum_n I_n \left\{ \cos\left(\omega_m \frac{\Delta}{2c}\right) \cos\left(\omega_n \frac{\Delta}{c}\right) \frac{2F}{\pi} \left[\phi_{n,-1} - 2\phi_{n,0} + \phi_{n,+1}\right] \right. \\
\left. + \sin\left(\omega_m \frac{\Delta}{2c}\right) \sin\left(\omega_n \frac{\Delta}{c}\right) \frac{2F}{\pi} \left[\phi_{n,-1} - \phi_{n,+1}\right] \right\} \\
+ J_0(\beta) J_1(\beta) \cos(\omega_m t) \times \\
\sum_n I_n \left\{ \cos\left(\omega_m \frac{\Delta}{2c}\right) \cos\left(\omega_n \frac{\Delta}{c}\right) \frac{2F}{\pi} \left[\delta_{n,-1} - \delta_{n,+1}\right] \right. \\
\left. - \sin\left(\omega_m \frac{\Delta}{2c}\right) \sin\left(\omega_n \frac{\Delta}{c}\right) \left[2 - \frac{2F}{\pi}(\delta_{n,-1} + 2\delta_{n,0} + \delta_{n,+1})\right] \right\}.
\end{aligned}
\qquad (6)
$$

Note that this expression reduces to the signal retrieved in cw NICE-OHMS when the OPD is equal to zero [28].

The NICE-OFCS interferogram resides on a DC offset originating from the beating of the comb lines with their own sidebands, as described above. This offset, in the weakly absorbing sample approximation, can be expressed as

$$I_{\omega_m}^{OFFSET} = J_0(\beta) J_1(\beta) \sin(\omega_m t) \times \\ \sum_n I_n \cos\left(\omega_m \frac{\Delta}{2c}\right) \frac{2F}{\pi} \left[\phi_{n,-1} - 2\phi_{n,0} + \phi_{n,+1}\right] \\ + J_0(\beta) J_1(\beta) \cos(\omega_m t) \times \\ \sum_n I_n \cos\left(\omega_m \frac{\Delta}{2c}\right) \frac{2F}{\pi} \left[\delta_{n,-1} - \delta_{n,+1}\right]. \quad (7)$$

It is important to note that this DC offset is zero in the absence of an absorber; moreover, it is noise-immune since the amplitude attenuations and phase shifts caused by the cavity cancel. This implies that the DC offset does not couple in any noise originating from the frequency-to-amplitude noise conversion, which is the key to the noise-immunity of the FTS-based NICE-OFCS technique.

The intriguing feature of FTS-based NICE-OFCS, as compared to cw NICE-OHMS, is the existence of the absorption term that violates the noise-immune condition [in the last line of Eqs. (5) and (6)]. This term is not zero in the absence of an absorber; it is also not noise-immune since the attenuations of a comb line and its sidebands combine with the same sign, hence contributions from the frequency-to-amplitude noise conversion add up. However, after summation over all comb lines, the interferogram originating from this term has a form of a short burst. This implies that the noise couples in only during a small fraction (typically <1%) of the OPD scan in the FTS, which makes this noise contribution negligible.

**4. Experimental setup and procedures**

The experimental setup is shown in Fig. 2. Our OFC source is an Er:fiber femtosecond laser (MenloSystems, FC1500-250-WG) with a repetition rate of 250 MHz that provides two low power outputs (20 mW) directly from the oscillator and a high power output (up to 500 mW) after a fiber amplifier, all in the 1.5–1.6 μm wavelength range. The output of the OFC is phase-modulated at two frequencies − $f_{PDH}$ and $f_m$ − using a single fiber-coupled lithium-niobate electro-optic modulator (EOM, Thorlabs, LN65S-FC). The lower frequency, $f_{PDH}$, provided by a signal generator, is equal to 20 MHz and used for generating the sidebands for the two-point Pound-Drever-Hall (PDH) locking of the OFC to the cavity [6]. The phase-modulated OFC light is coupled into free space and spatially mode-matched to the $TEM_{00}$ mode of the cavity. The combination of the waveplates and the polarizing beam splitter picks up the light reflected from the cavity. The reflected beam is dispersed by a grating and two different wavelength regions (defined as locking points) of the dispersed light are incident on two separate photodetectors. The outputs of the photodetectors are synchronously detected at $f_{PDH}$ to yield two PDH error signals [29]. The first error signal is fed into a proportional-integral (PI) controller (New Focus, LB1005) that acts on the injection current of the OFC pump diode, and controls the $f_{ceo}$ of the OFC with a bandwidth of 150 kHz. The second error signal is split into two paths and fed to two separate PI controllers that work in parallel to control the $f_{rep}$ of the OFC. The first controller (home-built) is affecting the OFC resonator length via an intracavity PZT, controlling the $f_{rep}$ with a low bandwidth (6 kHz) but a large range (5 kHz). The second one (New Focus, LB1005) is connected to an EOM inside the OFC resonator, which modulates the intracavity refractive index and thus changes the $f_{rep}$ with a high bandwidth (500 kHz) but a small range (0.1 Hz) [30].

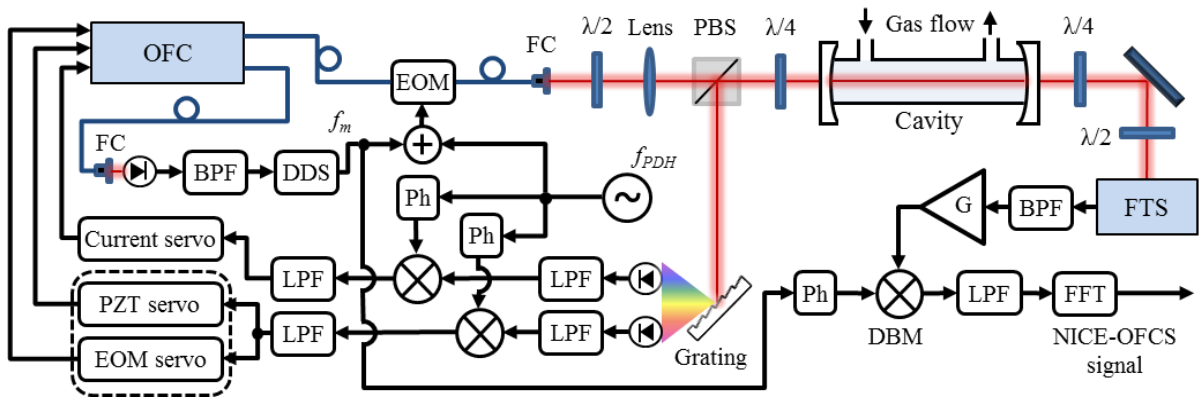

Fig. 2. Experimental setup: OFC – optical frequency comb; EOM – electro-optic modulator; FC – fiber collimator; λ/2 – half-waveplate; λ/4 – quarter-waveplate; PBS – polarizing beam splitter; FTS – Fourier transform spectrometer; BPF – band-pass filter; G – amplifier; LPF – low-pass filter; Ph – phase shifter; DBM – double-balanced mixer; FFT – fast Fourier transform; DDS – direct digital synthesizer; $f_{PDH}$ – Pound-Drever-Hall modulation frequency; $f_m$ – NICE-OFCS modulation frequency.

We use two cavities with different lengths and finesses; a 45 cm long cavity (cavity 1, $FSR_1$ of 333.3 MHz) with a finesse of ~2300, and an 80 cm long cavity (cavity 2, $FSR_2$ of 187.5 MHz) with a finesse of ~9000. Both cavities are made of pairs of identical mirrors (Layertec) with a 5 m radius of curvature. The comb-cavity matching for these cavities is shown in Fig. 1(b) and (c), respectively. Cavity 1 yields 1 GHz spacing between the transmitted OFC lines. For this cavity, we use a modulation frequency, $f_{m1}$, equal to the cavity $FSR_1$, i.e. 333.3 MHz. Cavity 2 yields 750 MHz spacing between the transmitted OFC lines. For practical reason (availability of the RF components) for this cavity we use a modulation frequency, $f_{m2}$, equal to three times the cavity $FSR_2$, i.e. 562.5 MHz. Because of the different losses in the cavity mirrors, cavity 1 transmits 66% of the incident power per comb line, while cavity 2 transmits only 12%. This implies that although cavity 2 transmits more comb lines than cavity 1, the average transmission per nanometer of spectral bandwidth is a quarter of that of cavity 1. In order to compensate for this difference in transmission we use the oscillator output of the OFC with cavity 1 and the amplifier output with cavity 2. We adjust the amplifier power in front of the EOM using a variable fiber attenuator (OZ Optics) so that the total optical power transmitted through cavity 2 is the same as the power transmitted through cavity 1 using the oscillator output. The different parameters of both cavities are summarized in Table 1.

We target the $3\nu_1$ overtone band of $CO_2$ around 1575 nm, as it coincides with the maximum power of the oscillator output. When using the amplifier, we adjust its spectral shape so that it also has a maximum around this wavelength. We choose the locking points at 1572 nm for the current lock and at 1579 nm for the PZT+EOM lock in order to transmit only the fraction of the incident spectrum that overlaps with the $CO_2$ absorption band. Due to this choice of locking points, 70% of the spectral range of the oscillator and 80% of the spectral range of the amplifier output is transmitted through cavity 1 and 2, respectively. The resulting average power in cavity transmission is 1.4 mW for both cavities.

The cavity mirrors are mounted on stainless steel spacer tubes connected to a gas system in a flow configuration, which allows filling the cavities with gas samples at a desired pressure. The available gases are pure $N_2$, 1% $CO_2$ in $N_2$, and 1000 ppm $CO_2$ in $N_2$, the latter two specified with an accuracy of 1%. The flows of the buffer gas ($N_2$) and the $CO_2$ standards are adjusted independently with programmable flow controllers (Bronkhorst, F-201CV). This allows mixing the $CO_2$ standards with pure $N_2$ to obtain lower $CO_2$ concentrations in the cavity. A vacuum pump is connected to the cavity outlet via a programmable pressure controller (Bronkhorst, P-702CV), which stabilizes the pressure inside the cavity.

Table 1. The summary of the properties of the two cavities used in this work and the achieved performance.

| Cavity | 1 | 2 | Units |
|---|---|---|---|
| Length (L) | 45 | 80 | cm |
| Free spectral range (FSR) | 333.3 | 187.5 | MHz |
| Finesse (F) @ 1579 nm | 2150 | 9400 | --- |
| Modulation frequency ($f_m$) | 333.3 | 562.5 | MHz |
| Transmitted comb line spacing | 1 | 0.75 | GHz |
| Transmitted comb lines | 25 | 33 | % |
| Transmission per comb line | 66 | 12 | % |
| Transmitted bandwidth | 70 | 80 | % |
| Total average transmitted power | 11.5 | 3.2 | % |
| Noise equivalent absorption @ 1579 nm | $3.5 \times 10^{-8}$ | $4.2 \times 10^{-9}$ | $cm^{-1}\ Hz^{-1/2}$ |
| Sensitivity per spectral element | $6.2 \times 10^{-10}$ | $6.4 \times 10^{-11}$ | $cm^{-1}\ Hz^{-1/2}$ |
| $CO_2$ concentration detection limit | 3.3 | 0.45 | ppm $Hz^{-1/2}$ |

In the first demonstration of NICE-OFCS [25] we used a free-running tuneable RF source to generate the modulation frequency and manually tuned it to the proper value by minimizing the noise in the NICE-OFCS signal. Since the RF source frequency was not locked to the cavity FSR, the perfect matching between the two frequencies was lost after some time due to the drift of the cavity length. In order to compensate for this drift, we are now generating the modulation frequency with a direct digital synthesizer (DDS, Analog Devices, AD9915) using the 5[th] harmonic of the OFC repetition rate (i.e. 1.25 GHz) as a clock input. The clock signal is measured at the second oscillator output of the OFC with a fast photodetector (Electro-Optics Technology, 3000A), whose output is filtered with a band-pass filter to remove the other harmonics. In this configuration, the modulation frequency is locked to the repetition rate of the laser, which in turn is locked to the cavity FSR. This eliminates the need of manual tuning or long term control of the modulation frequency to compensate the drifts in the cavity length and enhances the long term stability of the system.

The light transmitted through the cavity is injected into a fast-scanning FTS, whose design is similar to that described in [19]. Two 2-inch diameter gold-coated retro-reflectors (Edmund optics) mounted back-to-back on a miniature linear translation stage (Parker, LX80L) constitute the ends of the moving arms of the interferometer. The scanning speed of the retro-reflectors is 0.2 m/s, which implies that the OPD between the interferometer arms is scanned at 0.8 m/s. The OPD is calibrated using a stabilized He-Ne laser (stability on the order of $10^{-7}$) whose beam is propagating parallel to the comb beam in the FTS. The OFC output of the interferometer is detected by a 1 GHz bandwidth InGaAs detector (Electro-Optics Technology, 3000A), while the He-Ne output is detected by a 2 MHz bandwidth Si detector (Thorlabs, PDA36A-EC). The average comb power reaching the InGaAs detector is 0.37 mW for both cavities. The output of this detector is synchronously demodulated at the modulation frequency ($f_{m1}$ or $f_{m2}$) at a phase adjusted using a phase shifter between the DDS output and the local-oscillator input of the double-balanced mixer (DBM). The He-Ne interferogram and the demodulated comb interferogram are recorded with a 2-channel data acquisition card (National Instruments, PCI-5922) at 5 Msample/s and 20-bit resolution. The comb interferogram is resampled at the zero-crossings and extrema of the He-Ne interferogram to yield an OPD-calibrated NICE-OFCS interferogram. Fast Fourier transform of the NICE-OFCS interferogram yields finally a NICE-OFCS signal.

## 5. Results

### 5.1. NICE-OFCS interferogram

The magnitude of the different absorption and dispersion terms in the FTS-based NICE-OFCS signal depends on the ratio of the modulation frequency to the linewidth of the probed transitions, as is the case in ordinary FMS [27] or NICE-OHMS [31]. Both dispersion terms present in the in-phase interferogram, as well as the first absorption term present in the out-of-phase interferogram [see Eq. (6)] are maximized when the modulation frequency is roughly equal to the full-width-at-half-maximum (FWHM) linewidth of the probed absorption lines. The second absorption term, however, has a different dependence on the ratio of $f_m$ to the linewidth, since the attenuations of a comb line and its sidebands combine with the same sign. In contrast to the other three terms, this signal is large also in the undermodulated case, i.e. when the modulation frequency, $f_m$, is smaller than the FWHM linewidth of the absorption line. Moreover, this signal contains a background, which enables normalization and removes the need for calibration. These two features make detection of the out-of-phase interferogram the preferred mode of operation.

Figure 3 shows the out-of-phase NICE-OFCS interferograms from both cavities recorded for an OPD scan from 0 to 120 cm, which lasts 1.5 s. The dominating features in the out-of-phase interferogram are the bursts originating from the background signal present in the last term of Eqs. (5) and (6). Due to the different effective comb line spacing in cavity transmission and different modulation frequencies, the positions of these bursts and their relative amplitudes are dissimilar for the two cavities. In general, the amplitudes of the bursts follow the sine envelope [the envelopes from Eqs. (5) and (6) are drawn by the dashed lines in the figure as guidance to the eye], wherefore there are no bursts where this envelope is equal to zero. The small deviations from the amplitudes determined by the sine envelope are caused by the divergence of the OFC beam in the FTS, which decreases the interferogram amplitude for larger OPDs.

For detection of the NICE-OFCS signal we limit the acquisition to a symmetric range around one burst (indicated in Fig. 3 for both cavities), which yields spectral resolution equal to the effective comb line spacing in cavity transmission. Such resolution is sufficient as long as the comb line spacing is lower than the linewidth of the absorption lines. As mentioned above, of the two molecular absorption terms in the out-of-phase interferogram, the one multiplied by the sine envelope [last line in Eqs. (5) and (6)] dominates over the one multiplied by the cosine envelope [second-last line in Eqs. (5) and (6)] in the undermodulated case. In this case, detecting the burst at OPD = 30 cm or 60 cm for cavity 1 yields the same NICE-OFCS signal, since both bursts are symmetrically placed with respect to the sine envelope and the contribution from the signal multiplied by the cosine envelope can be neglected. For cavity 2 we detect the burst at OPD = 80 cm, because it has the largest magnitude and is symmetric with respect to both envelopes. The scan over the indicated acquisition ranges takes 0.38 and 0.50 s for the two cavities, respectively. We acquire the signal only during one direction of the scan of the FTS retro-reflectors, which imposes dead time in repeated measurements that is used for processing and saving the data. As a result, for cavity 1 and 2 one interferogram is acquired every 1 s and every 1.6 s, respectively.

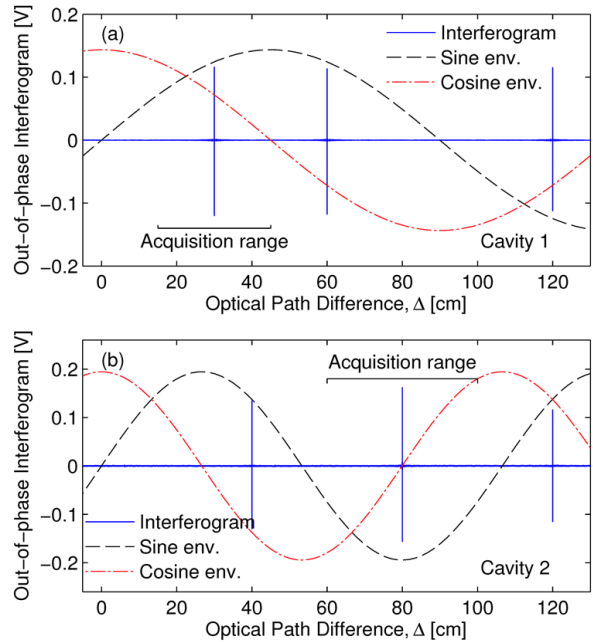

Fig. 3. The out-of-phase NICE-OFCS interferogram as a function of optical path difference (blue) together with the sine (dashed black) and cosine (dashed–dotted red) envelope factors, drawn as a guidance to the eye, for (a) cavity 1 ($L = 45$ cm) and (b) cavity 2 ($L = 80$ cm).

## *5.2. NICE-OFCS signal*

Figure 4 shows $CO_2$ NICE-OFCS absorption signals from the two cavities normalized to corresponding background spectra taken with cavities filled with pure nitrogen (black curves, 20 averages). Cavity 1 was filled with 1% of $CO_2$ in $N_2$, and cavity 2 with 200 ppm of $CO_2$ in $N_2$ (obtained by diluting the 1000 ppm $CO_2$ mixture with pure $N_2$), both at a total pressure of 500 Torr. At this pressure, the average FWHM linewidth of the $CO_2$ absorption lines is 3 GHz.

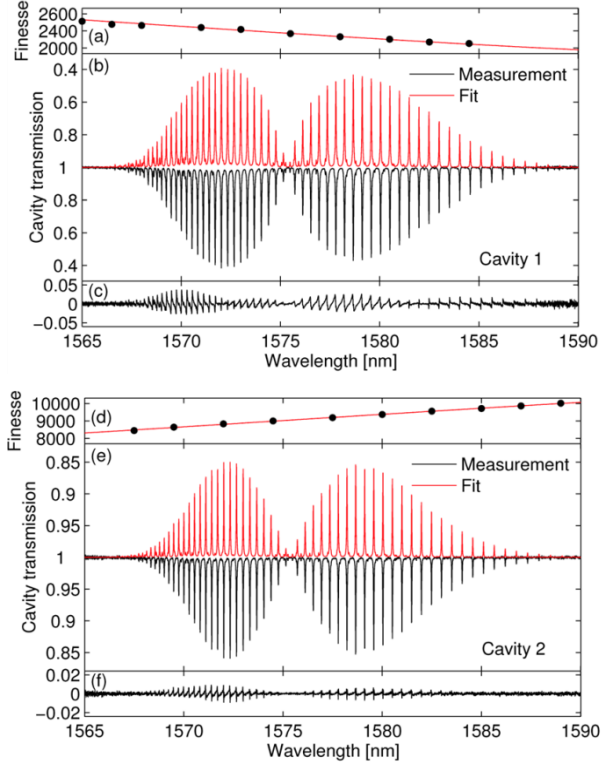

Fig. 4. Experimental and theoretical NICE-OFCS signals from cavity 1 (upper figure) and 2 (lower figure). (a) and (d) Cavity finesse measured by cavity ring-down (black markers) along with a 3$^{rd}$ order polynomial fit (red). (b) and (e) Normalized NICE-OFCS absorption signals (black, 20 averages) from (b) 1% $CO_2$ in $N_2$ in cavity 1 and (e) 200 ppm of $CO_2$ in $N_2$ in cavity 2, both at 500 Torr total pressure, along with fitted spectra calculated using the model described in the text (red, inverted for clarity). (c) and (f) Residuals of the fits.

We compare the experimental absorption spectra to the theoretical model by performing a fit to the data. We calculate the theoretical spectrum using the last factor in the last line of Eq. (5), i.e. $\mathrm{Re}\left(T_{n,0}T_{n,-1}^* + T_{n,0}^*T_{n,+1}\right)$, hence neglecting the term in the previous line, which is justified in the undermodulated case (the FWHM linewidth of the absorption lines is much broader than the modulation frequencies used). We also neglect the contribution from the slowly varying sine envelope of the interferogram. We calculate the attenuation and phase shift caused by the individual absorption lines using spectral line parameters from the HITRAN database [32] and a complex Voigt profile under the prevailing experimental conditions. We neglect the frequency offset between the comb lines and cavity modes, i.e. we put the cavity phase shift, $\varphi_{n,k}$, to an integer multiple of $2\pi$. We also take into account the wavelength dependence of the cavity finesse, which was measured by cavity ring-down. The measured values of the finesse are shown by black markers in Fig. 4(a) and (d) together with a 3$^{rd}$ order polynomial fit to the data (red) for the two cavities, respectively. The theoretical spectrum is fitted to the experimental data with $CO_2$ concentration as a fitting parameter, together with a 3$^{rd}$ order polynomial and a sum of low frequency sinusoidal etalon fringes to correct the baseline. The experimental data (after baseline correction) and the fitted $CO_2$ spectra (inverted for clarity) are shown in black and red, respectively, in Fig. 4(b) and (e), with the corresponding residuals in the lower panels, (c) and (f). As can be seen from the figure, the theoretical model and the experimental data show good agreement over the entire spectral range.

The main reason for the dissimilar structure in the residuals of the fits to the signals measured from the two cavities is that the two cavities are made with mirrors from different coating runs. Mirrors for cavity 1 were designed for minimum dispersion at 1530 nm, while those for cavity 2 – for 1570 nm. This implies that the spectra from cavity 1 are measured away from the design wavelength, which causes a slight walk-off between the comb lines and cavity modes and thus dispersion overshoots in the absorption lines [6]. This frequency offset between the comb lines and cavity modes was not included in the fitting model, which is the cause of the asymmetric structures visible in the residuum. The spectra from cavity 2 are affected by the dispersion of the cavity mirrors to a much smaller extent, since they are measured close to the design wavelength of the mirrors. The structure visible in the residuum of the spectrum measured in cavity 2 is more symmetric around each absorption line and indicates that the fitted model needs to be expanded to include the influence of the sine envelope as well as the absorption term multiplied by the cosine envelope factor. However, even using this simplified model, the $CO_2$ concentrations returned by the fits were 0.99% for cavity 1 and 210 ppm for cavity 2. The first value is within the specified 1% uncertainty of the $CO_2$ sample; the larger discrepancy of the second value (5%) can partly be explained by inaccuracies in the gas mixing process.

*5.3. Sensitivity*

To estimate the absorption sensitivity we take the ratio of two consecutive background spectra and fit and remove a baseline as in the treatment of the absorption signals above for both cavities. We evaluate the noise at 1579 nm, which corresponds to the center of the P branch of the $CO_2$ absorption band. The standard deviation of the noise is found to be $\sigma_1 = 2.5 \times 10^{-3}$ at 0.76 s for cavity 1, and $\sigma_2 = 2.0 \times 10^{-3}$ at 1 s for cavity 2. This corresponds to a noise equivalent absorption coefficient of $3.5 \times 10^{-8}$ cm$^{-1}$ Hz$^{-1/2}$ and $4.2 \times 10^{-9}$ cm$^{-1}$ Hz$^{-1/2}$ for the two cavities, respectively, calculated as $\sigma T^{1/2} / L_{eff}$, where $T$ is the acquisition time of two interferograms and the effective length $L_{eff}$ is defined as $2FL/\pi$, with finesse equal to 2150 and 9400 at 1579 nm for the two cavities, respectively. The number of resolved elements, $M$, in the spectral range shown in Fig. 4 is 3200 for cavity 1 and 4300 for cavity 2. Thus, the absorption sensitivity per spectral element, calculated as $\sigma / L_{eff} (T/M)^{1/2}$, is equal to $6.2 \times 10^{-10}$ cm$^{-1}$ Hz$^{-1/2}$ for cavity 1 and $6.4 \times 10^{-11}$ cm$^{-1}$ Hz$^{-1/2}$ for cavity 2.

By calculating the $CO_2$ concentration that yields a NICE-OFCS absorption signal with amplitude equal to the standard deviation of the noise on the baseline, $\sigma$, for the strongest absorption line in the P branch we find that the noise equivalent $CO_2$ concentration limit is equal to 22 ppm and 2.5 ppm of $CO_2$ at 1 s integration time for cavity 1 and 2, respectively. A better way to estimate the concentration detection limit is to make use of the entire spectrum and the multiline fitting advantage [22]. To do this, we measure background spectra with cavities filled with pure nitrogen at 500 Torr for one hour and normalize them to the first spectrum. Afterwards we fit a sum of the simplified NICE-OFCS $CO_2$ spectrum and a baseline to these normalized spectra. The Allan-Werle plot [33] of the concentrations found from these fits is shown in Fig. 5 for cavity 1 (solid blue markers) and 2 (solid red markers). The two dashed black lines show the $\tau^{-1/2}$ dependence characteristic for the white-noise dominated regime that is fitted to the corresponding measurement points. The $CO_2$ concentration detection limits, given by the slopes of the fitted lines, are 3.3 ppm $Hz^{-1/2}$ for cavity 1 and 0.45 ppm $Hz^{-1/2}$ for cavity 2. Note that these detection limits are a factor of ~6 lower than the noise equivalent $CO_2$ concentration calculated for a single absorption line in the center of the P branch. The fact that the detection limit for cavity 2 is ~7 times lower than for cavity 1 agrees well with the ratio of the products of the finesse and length of the two cavities, which determine the absorption path length enhancement [(9000×80)/(2300×45) ≈ 7]. The Allan-Werle plot shows that the absolute minimum detectable concentration of $CO_2$ is 140 ppb after 520 s of integration in cavity 1, and 25 ppb after 330 s in cavity 2.

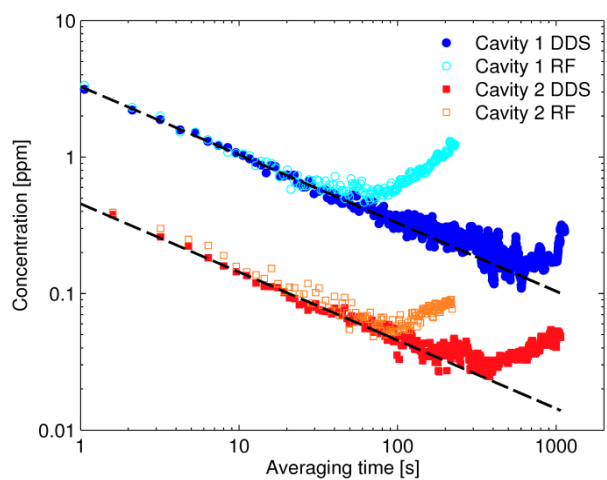

Fig. 5. The Allan-Werle plot of the minimum detectable $CO_2$ concentration retrieved from fitting of NICE-OFCS $CO_2$ spectra to normalized background spectra measured in the two cavities using different modulation frequency sources (cavity 1 with DDS - solid blue circular markers, cavity 1 with RF - open cyan circular markers, cavity 2 with DDS - solid red square markers, and cavity 2 with RF source - open orange square markers). The dashed black lines show the linear fits to the white-noise dominated regimes of each cavity measurement using the DDS.

To demonstrate the long-term stability improvement when using the $f_{rep}$-referenced DDS instead of the fixed frequency RF source for generating the modulation frequency, we show in Fig. 5 for comparison the concentrations retrieved from fits to normalized background spectra measured using the RF source and cavity 1 (open cyan markers) and 2 (open orange markers). For short integration times, using the DDS and the RF source yields almost the same noise levels since the RF source is initially tuned precisely to the correct frequency. At longer times, however, the cavity FSR drifts due to the thermal expansion of the cavity length causing a mismatch between the fixed RF frequency and the cavity FSR. This mismatch deteriorates the noise immunity and causes increased noise in the spectrum and thus worse detection limits. This is reflected in the linear drift visible in the Allan-Werle plot after 60 s for cavity 1 and after 90 s for cavity 2.

## 6. Conclusions

Fourier-transform-based NICE-OFCS is a broadband, highly sensitive, and high-resolution spectroscopic technique that allows acquiring entire spectral bands in acquisition times on the order of a second. Thanks to the immunity of the technique to the frequency-to-amplitude noise conversion, the detection sensitivity is directly scalable with the cavity finesse. The NICE-OFCS detection scheme can be implemented in commercial FTIR instruments using a single high-speed detector at the interferometer output and standard RF components for phase-sensitive detection. The various possible comb-cavity matching solutions provide flexibility in the choice of cavity lengths and modulation frequencies for a given OFC with a fixed repetition rate. The long term stability and low-noise performance of the system requires a perfect match of the modulation frequency to the cavity FSR. We have shown that this match can be passively sustained by a simple approach in which the modulation frequency is generated by a DDS referenced to the repetition rate of the OFC (and thus to the cavity FSR due to the comb-cavity lock) instead of a fixed frequency source. We achieved an absorption sensitivity of $6.4 \times 10^{-11}$ cm$^{-1}$ Hz$^{-1/2}$ per spectral element in a cavity with a finesse of ~9000, and a minimum detectable concentration of 25 ppb of $CO_2$ in 330 s, measured at the $3\nu_1$ overtone band around 1575 nm.

The functional form of the FTS-based NICE-OFCS signal differs from that familiar from cw NICE-OHMS. In particular, there exists a term in which the molecular signals have a shape similar to direct cavity-enhanced absorption on top of a background signal, which makes the NICE-OFCS technique calibration-free (provided the cavity finesse is known). We have shown that by using a simplified form of the full model, given by this absorption term, one can extract the concentration from the measured spectra with <5% error. Further improvements to the fitted model will lead to a better precision in concentration retrieval. Another step towards wider applicability of the NICE-OFCS technique is improving the spectral resolution by resolving the comb lines in order to allow for measurements of narrower absorption lines at lower pressures. The ability to fit a model to the experimental data provides also the possibility to utilize multiline fitting to obtain lower concentration detection limits compared to those calculated from the signal-to-noise ratio of a single absorption line in the spectrum.

**Acknowledgements**

This project was supported by the Swedish Research Council (621-2012-3650), Swedish Foundation for Strategic Research (ICA12-0031), the Carl Trygger's Foundation (CTS12:131), and the Faculty of Science and Technology, Umeå University. The authors thank Piotr Maslowski, Ticijana Ban, and Ove Axner for useful discussions about the NICE-OFCS principles.